\documentclass[sigconf,11pt]{acmart}

\renewcommand\footnotetextcopyrightpermission[1]{}
\usepackage{listings}
\usepackage{xcolor}
\usepackage{float}
\usepackage{booktabs}
\usepackage{graphicx}
\usepackage{seqsplit}
\usepackage{xurl}

\makeatletter
\def\@ACM@printacmbadges{}
\def\@acmBadgeL@image{}
\def\@acmBadgeR@image{}
\makeatother

\AtBeginDocument{%
  \providecommand\BibTeX{{%
    \normalfont B\kern-0.5em{\scshape i\kern-0.25em b}\kern-0.8em\TeX}}}

\setcopyright{none}
\copyrightyear{}
\acmYear{}
\acmDOI{}

\acmConference[CSE6242]{Make sure to enter the correct
  conference title from your rights confirmation emai}{Sept 01,
  2022}{Atlanta, GA}
\acmPrice{}
\acmISBN{}

\begin{document}

\title{\Large Bayesian Negative Binomial Regression of Afrobeats Chart Persistence}

\author{\normalsize Cabansag, Ian Jacob and Ntegeka, Paul}
\renewcommand{\shortauthors}{Cabansag and Ntegeka}



\maketitle 
\pagestyle{plain} 
\section{Abstract}

Afrobeats tracks compete for listener attention on streaming platforms, where chart visibility can translate directly into revenue and cultural impact. 
This project studies the relationship between collaboration and chart persistence using the 2024 Spotify Top 200 daily charts for Nigeria. 
Each track is summarized by the number of days it appears on the Top 200 over the year and its total annual streams in the Nigerian market. 
A Bayesian Negative Binomial regression model is fitted in which the response is the count of days on chart and the predictors are an indicator for collaboration (solo versus multi--artist track) and the log of total annual streams. 
This model appropriately handles over--dispersed count data and allows a direct interpretation of collaboration effects after controlling for overall popularity. 
Posterior inference is carried out in \texttt{PyMC} using NUTS, and posterior rate ratios, the posterior probability that collaboration helps, and posterior predictive checks are reported. 
The results suggest that, after conditioning on total streams, collaboration tracks have slightly \emph{fewer} days on the chart than comparable solo tracks, with an estimated rate ratio of approximately $\exp(\beta_{\text{collab}}) \approx 0.93$ (95\% interval $[0.88, 0.99]$) and posterior probability $P(\beta_{\text{collab}} > 0 \mid \text{data}) \approx 0.007$.

\section{Introduction}

The global rise of Afrobeats has been driven in part by streaming platforms, where editorial playlists and algorithmic recommendations amplify the visibility of hit songs. 
For artists and labels, one key indicator of success is \emph{chart persistence}: how long a track remains on a country's Top 200 chart. 
Longer persistence implies sustained listener interest and more opportunities for discovery.

A common belief among artists and industry observers is that collaboration for example, featuring another artist on a track helps a song stay relevant for longer. 
However, collaboration is also correlated with other factors such as marketing budgets and artist popularity. 
Simply comparing the average number of days on chart for solo and collaboration tracks confounds these effects.

The objective of this project is to quantify the effect of collaboration on chart persistence in a principled statistical framework. 
Daily Spotify Top--200 data for the Nigeria market in 2024 are used, focusing on Afrobeats and related genres as reflected in the chart. 
The central question is:

\medskip
\begin{quote}
\emph{After controlling for overall popularity (total annual streams), do collaboration tracks stay longer on the Nigeria Top 200 chart than solo tracks?}
\end{quote}
\medskip

From a methodological standpoint, the project moves beyond a simple conjugate Bayesian A/B test and instead employs a Bayesian Generalized Linear Model (GLM) with a Negative Binomial likelihood. 
This choice is motivated by the count nature and strong over--dispersion of the response variable (days on chart). 
The model is estimated using Markov Chain Monte Carlo (MCMC) with the NUTS algorithm implemented in \texttt{PyMC}, providing full posterior distributions over parameters and derived quantities.

The Data section describes the dataset and preprocessing. 
The Bayesian model section presents the Negative Binomial regression formulation and prior choices. 
The Computation section summarizes the computational setup and convergence diagnostics. 
The Results section reports posterior summaries and predictive checks, and the Discussion section concludes with a substantive interpretation of the findings and limitations.

\section{Data}

\subsection{Source and structure}
The dataset consists of daily snapshots of the Spotify Top--200 for the Nigeria market over the period 1 January 2024 to 31 December 2024. 
Each row corresponds to one track on one day and includes, among others, the following variables:
\begin{itemize}
    \item \texttt{date}: calendar date of the chart entry.
    \item \texttt{uri}: unique Spotify track identifier.
    \item \texttt{rank}: chart position (1--200) on that date.
    \item \texttt{track\_name}: track title.
    \item \texttt{artist\_names}: comma--separated list of artist names.
    \item \texttt{streams}: number of streams for that track in Nigeria on that date.
\end{itemize}

The data were obtained from the official Spotify Charts website for the Nigeria regional daily chart \citep{spotifycharts}:
\begin{center}
\url{https://charts.spotify.com/charts/view/regional-ng-daily/latest}
\end{center}

The raw dataset contains $N = 72{,}995$ daily track--level observations, representing $n_{\text{tracks}} = 1{,}335$ unique tracks. 
The date range covers the full year, from 2024--01--01 to 2024--12--31.

\subsection{Track--level aggregation}

The regression analysis is conducted at the track level, not the day level. 
For each track $i$ (identified by a unique URI), the following summary variables are constructed:
\begin{itemize}
    \item $n_i$: the number of distinct days in 2024 on which track $i$ appears in the Nigeria Top--200. This is referred to as \emph{days on chart}.
    \item $S_i$: the total number of streams accumulated by track $i$ across all days in which it appears on the chart, $S_i = \sum_{t} \text{streams}_{it}$.
    \item $x_{i,\text{collab}}$: an indicator for whether track $i$ is a collaboration ($1$) or a solo track ($0$).
\end{itemize}

The collaboration indicator is derived from the \texttt{\seqsplit{artist\_names}} field using a simple text heuristic.
Tracks are classified as collaborations if the artist field contains a comma (multiple artists), an ampersand ``\&'', or tokens such as ``feat'', ``ft.'' or `` x '' indicating a featured artist. 
This heuristic agrees well with manual inspection of a random subset of tracks.

For numerical stability and interpretability in the regression, the log--transformed total streams,
\[
x_{i,\text{logStreams}} = \log S_i,
\]
are used as a predictor.

Table~\ref{tab:summary} reports summary statistics for the main variables across the $n_{\text{tracks}} = 1{,}335$ tracks.

\begin{table}[t]
\centering
\caption{Summary statistics for track-level variables (Nigeria Spotify Top 200, 2024).}
\label{tab:summary}
\resizebox{\columnwidth}{!}{%
\begin{tabular}{lcccc}
\toprule
Variable & Mean & SD & Min & Max \\
\midrule
Days on chart $n_i$ & 54.7 & 81.9 & 1.0 & 365.0 \\
Total streams $S_i$ & 1,908,473 & 3,532,364 & 10,132 & 29,251,917 \\
$\log S_i$ & 12.735 & 2.100 & 9.223 & 17.191 \\
Collaboration rate $P(x_{i,\text{collab}}=1)$ & \multicolumn{4}{c}{0.456} \\
\bottomrule
\end{tabular}}
\end{table}

Exploratory plots (histogram of $n_i$, boxplot of $n_i$ by collaboration status, and scatterplot of $n_i$ against $\log S_i$) show that the response variable is strongly right skewed and over dispersed relative to a Poisson distribution, and that both collaboration status and total streams are associated with days on chart. 
These features motivate the use of a Negative Binomial regression model in the next section.

\section{Bayesian Model}
\subsection{Likelihood: Negative Binomial regression}

Let $n_i$ denote the number of days track $i$ appears on the Top 200 chart in 2024. 
These are non-negative counts with substantial over-dispersion: $\operatorname{Var}(n_i)$ is considerably larger than $\mathbb{E}[n_i]$. 
The Negative Binomial distribution provides a flexible parametric family for such data.

For each track $i$ the model assumes
\begin{equation}
    n_i \mid \mu_i, \alpha \sim \text{NegBin}(\mu_i, \alpha),
\end{equation}
where $\mu_i > 0$ is the mean and $\alpha > 0$ is a dispersion parameter. 
Under this parameterization,
\[
\mathbb{E}[n_i \mid \mu_i, \alpha] = \mu_i, 
\qquad
\operatorname{Var}(n_i \mid \mu_i, \alpha) = \mu_i + \frac{\mu_i^2}{\alpha}.
\]

To relate $\mu_i$ to covariates, a log--link function with two predictors is used: collaboration status and log total streams. 
Let $x_{i1} = x_{i,\text{collab}}$ and $x_{i2} = x_{i,\text{logStreams}}$. 
The linear predictor is
\begin{equation}
    \eta_i = \beta_0 + \beta_1 x_{i1} + \beta_2 x_{i2},
\end{equation}
and the mean is linked via
\begin{equation}
    \mu_i = \exp(\eta_i).
\end{equation}

The coefficient $\beta_1$ captures the log--multiplicative effect of being a collaboration track, holding total streams fixed:
\[
\exp(\beta_1) = 
\frac{\text{Expected days on chart for collab track}}{\text{Expected days on chart for solo track}}
\quad \text{at fixed } S_i.
\]
Similarly, $\beta_2$ describes how days on chart scale with overall streams.

\subsection{Priors}

Weakly informative priors are placed on the regression coefficients and dispersion parameter. 
The coefficients $(\beta_0, \beta_1, \beta_2)$ receive independent Normal priors with mean zero and standard deviation $2$:
\begin{align}
    \beta_0 &\sim \mathcal{N}(0, 2^2), \\
    \beta_1 &\sim \mathcal{N}(0, 2^2), \\
    \beta_2 &\sim \mathcal{N}(0, 2^2).
\end{align}
These priors allow for substantial variation (for example, a priori $\exp(\beta_j)$ lies roughly between $e^{-4}$ and $e^{4}$ with high probability), while still gently regularizing extreme coefficients.

The dispersion parameter $\alpha$ is constrained to be positive and receives a Half--Normal prior:
\begin{equation}
    \alpha \sim \text{HalfNormal}(2).
\end{equation}
This prior favors moderate values of over--dispersion while allowing the data to inform the degree of extra--Poisson variability.

\subsection{Posterior and derived quantities}

Let $\boldsymbol{\beta} = (\beta_0, \beta_1, \beta_2)$ and denote all parameters by $\theta = (\boldsymbol{\beta}, \alpha)$. 
Given data $\mathbf{n} = (n_1,\dots,n_{n_{\text{tracks}}})$ and covariates $\mathbf{x}$, the posterior density is
\begin{equation}
    p(\theta \mid \mathbf{n}, \mathbf{x})
    \propto
    \left[\prod_{i=1}^{n_{\text{tracks}}}
        p\bigl(n_i \mid \mu_i(\theta), \alpha\bigr)
    \right]
    p(\boldsymbol{\beta}) p(\alpha).
\end{equation}

Two derived quantities are of particular interest:
\begin{enumerate}
    \item The collaboration \emph{rate ratio}
    \[
    R_{\text{collab}} = \exp(\beta_1),
    \]
    which measures, at fixed $S_i$, how many times more (or fewer) days a collaboration track is expected to stay on chart compared with a solo track.
    \item The posterior probability that collaboration helps:
    \[
    P(\beta_1 > 0 \mid \mathbf{n}, \mathbf{x}),
    \]
    which serves as a Bayesian measure of evidence for a positive collaboration effect.
\end{enumerate}

Posterior predictive distributions for $n_i$ are also computed and compared to the observed distribution as a model check.

\section{Computation}

Posterior inference is performed using the \texttt{PyMC} library (version~4.0.1) in Python, together with 
\texttt{ArviZ} (version~0.21.0) for diagnostics and visualization. The model is implemented using a 
\texttt{NegativeBinomial} likelihood with a mean--dispersion parameterization and a log--link. Sampling 
is carried out using the No-U-Turn Sampler (NUTS).

Two chains are run with $2{,}000$ warmup iterations and $2{,}000$ post--warmup draws each, yielding a total 
of $4{,}000$ posterior samples. The target acceptance rate is set to $0.9$ to encourage smaller step sizes 
and stable sampling.

Convergence is assessed using the $\widehat{R}$ diagnostic and effective sample sizes reported by 
\texttt{ArviZ}. All parameters have $\widehat{R}$ values very close to $1.00$ and effective sample sizes 
in the thousands, indicating good mixing. Trace plots for $\beta_1$, $\beta_2$, and $\alpha$ do not show 
evidence of nonstationarity or multimodality.

Posterior predictive samples are generated using the PyMC posterior predictive sampler
(\texttt{\nolinkurl{pm.sample_posterior_predictive}}), resulting in replicated draws $n_i^{\mathrm{rep}}$ for each track.

A Poisson regression model with identical covariates is also considered. Posterior predictive diagnostics 
show that the Poisson model underestimates the variance and fails to reproduce the heavy right tail of 
highly persistent tracks, whereas the Negative Binomial model provides a substantially better fit. This 
supports the Bayesian modeling framework described above.

\section{Results}

\subsection{Posterior summaries}

Table~\ref{tab:posterior} reports posterior means and 94\% highest density intervals (HDIs) for the regression coefficients and dispersion parameter, as produced by \texttt{ArviZ}. 

\begin{table}[ht]
    \centering
    \caption{Posterior summaries for model parameters (means and 94\% HDIs).}
    \label{tab:posterior}
    \begin{tabular}{lrrrr}
        \toprule
        Parameter & Mean & SD & 3\% HDI & 97\% HDI \\
        \midrule
        $\beta_0$ (intercept)        & $-8.126$ & $0.122$ & $-8.354$ & $-7.904$ \\
        $\beta_1$ (collaboration)    & $-0.074$ & $0.030$ & $-0.129$ & $-0.014$ \\
        $\beta_2$ (log streams)      & $0.859$  & $0.009$ & $0.843$  & $0.876$ \\
        $\alpha$ (dispersion)        & $5.037$  & $0.237$ & $4.621$  & $5.503$ \\
        \bottomrule
    \end{tabular}
\end{table}

The coefficient on log total streams, $\beta_2$, is strongly positive, indicating that tracks with higher total streams tend to stay on the chart longer, as expected. 
The main quantity of interest is $\beta_1$, the effect of collaboration. 
The posterior for $\beta_1$ is centered at $-0.074$ with a 94\% HDI from $-0.129$ to $-0.014$, lying entirely below zero.

From the posterior for $\beta_1$ the collaboration rate ratio
\[
R_{\text{collab}} = \exp(\beta_1)
\]
has approximate posterior median $R_{\text{collab}} \approx 0.929$ and 95\% credible interval $[0.879, 0.986]$. 
The posterior probability that the effect is positive is very small:
\[
P(\beta_1 > 0 \mid \mathbf{n}, \mathbf{x}) \approx 0.007.
\]
In words, after controlling for total annual streams, collaboration tracks are expected to have slightly \emph{fewer} days on the Top--200 chart than comparable solo tracks, and there is strong Bayesian evidence against a positive collaboration effect.

\subsection*{Expected days on chart for solo vs collaboration}

To provide a more interpretable summary, consider a ``typical'' track at the median log total streams, $\tilde{x}_{\text{logStreams}} \approx 12.71$. 
For each posterior draw of $(\beta_0,\beta_1,\beta_2)$, the expected days on chart for solo and collaboration tracks at this reference value are
\begin{align*}
    \mu_{\text{solo}}^{(s)} &= \exp\bigl(\beta_0^{(s)} + \beta_2^{(s)} \tilde{x}_{\text{logStreams}}\bigr), \\
    \mu_{\text{collab}}^{(s)} &= \exp\bigl(\beta_0^{(s)} + \beta_1^{(s)} + \beta_2^{(s)} \tilde{x}_{\text{logStreams}}\bigr).
\end{align*}

Using a log--normal approximation based on the posterior means and standard deviations, the expected days on chart for a representative solo track are approximately
$\mu_{\text{solo}} \approx 16.3$ (95\% interval about $[11.8, 22.7]$), whereas for a representative collaboration track the corresponding value is
$\mu_{\text{collab}} \approx 15.2$ (95\% interval about $[10.9, 21.2]$). 
These values reinforce the interpretation that collaboration is associated with a small but non--negligible reduction in chart persistence when overall popularity is held fixed.

\subsection*{Posterior predictive checks}

\begin{figure}[ht]
    \centering
    \includegraphics[width=1\linewidth]{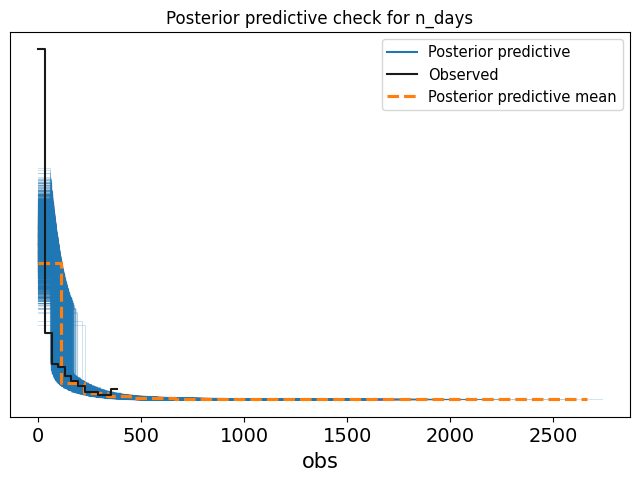}
    \caption{Posterior predictive check for days on chart. 
    The solid line represents the empirical density of $n_i$, while the shaded region shows the distribution of posterior predictive replicates $n_i^{\text{rep}}$.}
    \label{fig:ppc}
\end{figure}

Posterior predictive checks compare the empirical distribution of $n_i$ to that of replicated counts $n_i^{\text{rep}}$ drawn from the posterior predictive distribution. 
Figure~\ref{fig:ppc} shows a representative check, where the histogram (or kernel density estimate) of observed $n_i$ is overlaid with the distribution of posterior predictive replicates.

The Negative Binomial model reproduces the skewed distribution of days on chart reasonably well, including the long right tail of tracks with very high persistence, although it slightly underestimates the frequency of the most extremely persistent hits (e.g., tracks with more than 300 days on chart). 
Posterior predictive plots stratified by collaboration status (not shown) indicate that the model also captures the difference in distribution between solo and collaboration tracks.

\section*{Discussion}

This project applied a Bayesian Negative Binomial regression model to study chart persistence for Afrobeats tracks in the 2024 Nigeria Spotify Top--200. 
By aggregating daily chart data to track--level counts and modeling days on chart as a function of collaboration status and log total streams, the analysis moves beyond a simple A/B comparison and quantifies collaboration effects while controlling explicitly for overall popularity.

The main substantive finding is that collaboration is associated with a \emph{slight disadvantage} in terms of days on chart, once total annual streams are held fixed. 
The posterior rate ratio estimate of approximately $0.93$ implies that, at the same total stream level, collaborations are expected to have about $7\%$ fewer days on the chart than solo tracks. 
The posterior probability that collaboration has a beneficial effect is only about $0.7\%$, indicating strong Bayesian evidence against the notion that collaborations intrinsically extend chart lifetimes.

Methodologically, the project demonstrates a full Bayesian GLM workflow: formulation of an appropriate likelihood, specification of weakly informative priors, MCMC fitting with \texttt{PyMC}, interpretation of posterior effect sizes and probabilities, and posterior predictive model checking. 
In particular, the use of the Negative Binomial likelihood is essential to accommodate the over--dispersion in days on chart; a Poisson model underestimates variability and provides a noticeably poorer predictive fit.

Several limitations should be noted. 
First, the collaboration indicator is based on a heuristic parsing of artist names and may misclassify some tracks. 
Second, the model conditions on total annual streams, which are themselves the result of a complex process involving playlist placement, marketing, and recommendation algorithms. 
A more ambitious hierarchical model could attempt to jointly model streams and days on chart, or to introduce artist--level random effects. 
Third, the analysis is restricted to a single year and a single market; extending the study across years or countries could reveal whether the collaboration effect is stable over time and geography.

Despite these limitations, the model provides a clear demonstration of how Bayesian regression can be applied to real streaming data to address a concrete substantive question. 
Future work could build on this framework by adding genre information, label indicators, or artist network features to better understand the mechanisms through which collaboration influences chart performance.


\nocite{gelman2013,pymc,spotifycharts}
\bibliographystyle{ACM-Reference-Format}
\bibliography{references}

\appendix
\section{PyMC model code}
\noindent
The following Python code fragment shows the core \texttt{PyMC} model used in the analysis.

\begin{lstlisting}[language=Python, basicstyle=\ttfamily\small]
with pm.Model() as nb_model:
    beta0 = pm.Normal("beta0", 0.0, 2.0)
    beta_collab = pm.Normal("beta_collab", 0.0, 2.0)
    beta_log_streams = pm.Normal("beta_log_streams", 0.0, 2.0)

    alpha = pm.HalfNormal("alpha", 2.0)

    eta = beta0 + beta_collab * x_collab + beta_log_streams * x_log_streams
    mu = pm.math.exp(eta)

    obs = pm.NegativeBinomial("obs", mu=mu, alpha=alpha, observed=y)

    idata = pm.sample(draws=2000, tune=2000, chains=2, target_accept=0.9)
\end{lstlisting}

\end{document}